\documentclass[reprint,aps,pre,showpacs,twocolumn]{revtex4-2}

\usepackage{amsmath, amssymb}  
\usepackage{amsfonts} 
\usepackage{graphicx}
\usepackage{mathptmx, bm}
\usepackage{float}

\begin{document}

\title{Exact closed-form recurrence probabilities for biased random walks at any step number}

\author{Debendro Mookerjee$^{1}$}

\author{Sarah Kostinski$^{1}$}
\email{sk10775@nyu.edu}

\affiliation{\noindent \textit{$^{1}$Department of Physics, New York University, New York, NY 10003 USA}}

\date{\today}

\begin{abstract} 
\noindent

\noindent 

We report on a closed-form expression for the survival probability of a discrete 1D biased random walk to not return to its origin after $N$ steps.  Our expression is exact for any $N$, including the elusive intermediate range, thereby allowing one to study its convergence to the large $N$ limit. In that limit we recover Poly{\'a}'s recurrence probability, i.e. the survival probability equals the magnitude of the bias. We then obtain a closed-form expression for the probability of last return.  In contrast to the bimodal behavior for the unbiased case, we show that the probability of last return decays monotonically throughout the walk beyond a critical bias.  We obtain a simple expression for the critical bias as a function of the walk length, and show that it saturates at $1/\sqrt{3}$ for infinitely long walks. 
This property is missed when using expressions developed for the large $N$ limit. Finally, we discuss application to molecular motors' biased random walks along microtubules, which are of intermediate step number.

\vspace{5mm}

\end{abstract}

\maketitle

Results for 1D random walks are well-known in the limit of large step number, or long times. For example, consider Poly{\'a}'s celebrated result on the recurrence, i.e. return to the origin, of unbiased (symmetric) random walks: In the limit of long times, a symmetric random walk in dimensions $d \leq 2$ will always revisit its origin~\cite{Polya, Electric, Redner, Klafter}.  A key quantity in Poly{\'a}'s result, and more generally in Brownian motion, is the survival probability, which is the probability to not have reached a target for a given duration.  It is also the complement of the cumulative of the first passage time distribution, i.e. the probability distribution of times to first reach a designated target.  However, exact analytic expressions for such probabilities at short and intermediate times are lacking, in contrast to the long time limit.  To quote from a recent article~\cite{MajumdarWedge}: ``In the past, special attention has been devoted to the analysis of the large-time behavior of the survival probability, which displays a power-law decay... On the other hand, the analysis of the short-time behavior seems to have been left aside.''  The latter is our focus for the discrete 1D biased random walk.

 Survival probabilities at intermediate step numbers, or finite times, are important in certain contexts.  Consider the biological context of a molecular motor like dynein, which executes a 1D biased random walk on microtubules~\cite{Wordeman}.  Clearly, such molecules do not walk on microtubules for infinite time.  Obtaining a closed-form expression of the survival probability which holds for arbitrary step number, however, is not always tractable with the usual generating function approach. As Feller wrote in his classic text~\cite{Feller}: ``Given a generating function $P(s) = \sum p_k s^k$ the coefficients $p_k$ can be found by differentiations from the obvious formula $p_k = P^{(k)}(0)/k!$.  In practice it may be impossible to obtain explicit expressions and, anyhow, such expressions are frequently so complicated that reasonable approximations are preferable.''  Here we aim to show that an accessible closed-form expression for the survival probability (the complement of the recurrence probability) at arbitrary step number can be obtained using a path enumeration approach.  To the best of our knowledge, a closed-form expression for the survival probability, as given below in Eq.~\ref{S},
has not previously appeared in the literature.

To derive such an expression for the survival probability, consider a random walker that starts at the origin and jumps one step to the right or left on a lattice with respective probabilities $p$ and $q$.  Letting $+1$ ($-1$) correspond to a jump to the right (left), the walk can be represented as a sequence of $\pm 1$'s, e.g. $\{+1, +1, -1, +1, -1, -1, ...\}$. Thus the walker returns to the origin whenever the running sum is zero. 

Let us now consider the following two contributions to the survival probability: $R(N)$ is the probability that the walker remains to the right of the origin, and $L(N)$ is the probability that the walker remains to the left of the origin.  $R(N)$ thus accounts for walks whose running sum of the step sequence of $\pm 1$'s remains positive, and $L(N)$ accounts for those whose running sum remains negative.  The survival probability $S(N)$ is then simply $S(N) = R(N) + L(N)$.  Below we proceed to first obtain an expression for $R(N)$.

The probability that a particular sequence has $N_+$ steps to the right and $N_-$ steps to the left is $p^{N_+} q^{N_-}$. The total number of such sequences with $N = N_+ + N_-$ steps is $\binom{N}{N_-}$. Survival at the $N$th step means the random walker never returned to the origin up to step $N$, i.e. the running sum never reached $0$.  The fraction of such paths whose running sum remains above 0 is given by the ballot theorem~\cite{Renault,kostinski2016elementary}: $(N_+ - N_-)/N = 1 - 2 N_- / N$. We sum the probabilities of all such paths to obtain $R(N)$: 
\begin{align}
    R(N) &= \sum_{N_- = 0}^{\lfloor \frac{N-1}{2}\rfloor} p^{N-N_-}q^{N_-} \bigg(1 - \frac{2N_-}{N} \bigg) \binom{N}{N_-} 
\end{align}
where we have chosen the limits such that $N_+ > N_-$, or equivalently, $0 \le N_- \le \lfloor \frac{N-1}{2}\rfloor$.  This expression can be simplified to obtain (Supplementary Material~\cite{SupplMat}):
\begin{align}
    R&(N) = p^{N-\lfloor \frac{N-1}{2}\rfloor}q^{\lfloor \frac{N-1}{2}\rfloor} \binom{N-1}{\lfloor \frac{N-1}{2}\rfloor} \nonumber \\
    &+ \sum_{N_- = 0}^{\lfloor \frac{N-1}{2}\rfloor - 1}\bigg(p^{N - N_{-}}q^{N_{-}} - p^{N - N_{-} - 1}q^{N_{-} + 1}\bigg)\binom{N-1}{N_-} \, .
    \label{eq:Rprob}
\end{align}
The probability $L(N)$ that the running sum remains negative is the same as Eq.~\ref{eq:Rprob}, except that $p$ and $q$ are swapped, and $N_{-}$ is replaced by $N_{+}$. Adding the probabilities $R(N)$ and $L(N)$ yields the survival probability~\cite{SupplMat}:
\begin{align}
    S&(N) = \bigg( p^{\lceil \frac{N+1}{2}\rceil}q^{\lfloor \frac{N-1}{2}\rfloor} + p^{\lfloor \frac{N-1}{2}\rfloor}q^{\lceil \frac{N+1}{2}\rceil}\bigg)\binom{N-1}{\lfloor \frac{N-1}{2}\rfloor} \nonumber \\
    &+ (p-q)\sum_{x = 0}^{\lfloor \frac{N-1}{2}\rfloor - 1}\bigg(p^{N-x-1}q^x - p^xq^{N-x-1}\bigg)\binom{N-1}{x} \, .
    \label{comboSum}
\end{align}
Note the symmetry: interchanging $p$ and $q$ has no effect on the survival probability. Since $p+q=1$, terms in Eq.~\ref{comboSum} involving $p$ and $q$ can be expressed in terms of a single parameter, the bias $B=p-q$. An expression for the survival probability in terms of bias $B$ is thus obtained by replacing $p$ with $(1+B)/2$ and $q$ with $(1-B)/2$ in Eq.~\ref{comboSum}.  For small bias $|B| \ll 1$, a Taylor expansion reveals that $S(N,B) \approx c(N) + d(N) B^2$ where $c(N)$ is the symmetric walk result $S(N,0) = 2^{-(N-1)} \binom{N-1}{\lfloor \frac{N-1}{2} \rfloor}$ and $d(N)>0$.  In the limit of large bias $|B| \rightarrow 1$, a Taylor expansion shows that the survival probability approaches $|B|$.
\begin{figure}[t!]
\begin{centering}
\includegraphics[width=0.49\textwidth]{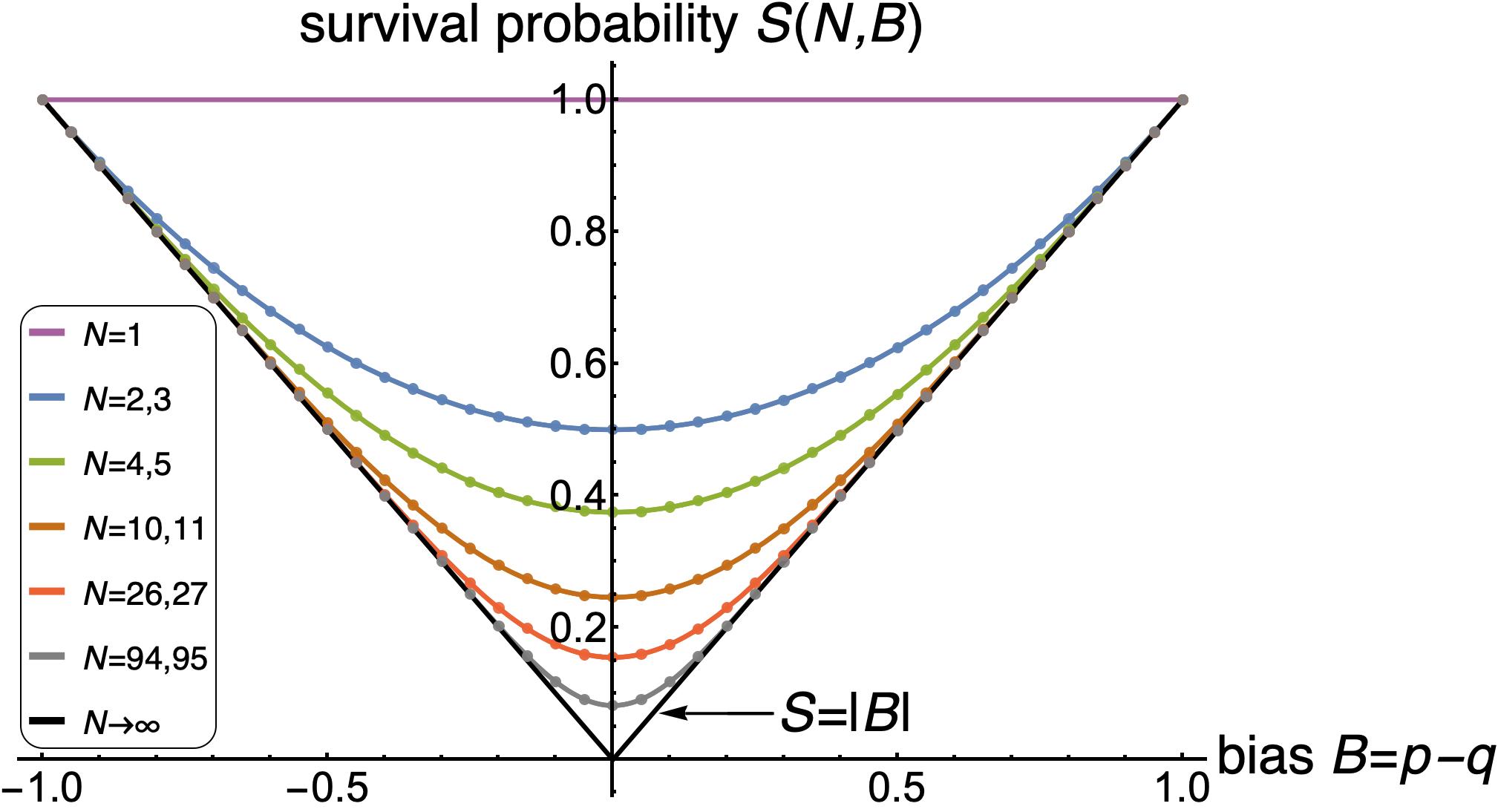}
\end{centering}
\vspace{-5mm}
\caption{Plot of the survival probability $S(N,B)$, i.e. the probability to never return to the origin through step $N$, versus the bias $B=p-q$. The colored curves correspond to our analytic result in Eq.~\ref{S} for different step numbers $N$. The colored points on those curves are simulation results (averaged over $10^6$ walks), which are in excellent agreement with Eq.~\ref{S}. As $N$ increases, $S$ approaches its large step number limit $S=|B|$ (illustrated in black). For small bias $|B| \ll 1$, the survival probability is quadratic in $B$.}
\label{Wedge}
\end{figure}

We find that the summation in Eq.~\ref{comboSum} can be expressed in terms of the Gauss hypergeometric function ${}_{2}F_{1}(a;b;c;z)$~\cite{Hypergeometric}. For the sake of brevity here, the derivation is provided in the Supplementary Material~\cite{SupplMat}.  To simplify the floor and ceiling functions, we separate the survival probability expression for odd and even step numbers ($S_{\text{odd}}$ and $S_{\text{even}}$, respectively):
\begin{subequations}
\label{S}
\begin{align}
     \label{oddS}
    &S_{\text{odd}}(N,B) = B + \frac{(1 - B^2)^\frac{N}{2}}{2^{N-1}} \binom{N-1}{\frac{N-1}{2}} \sqrt{\frac{1 - B}{1 + B}}  \bigg[1 \nonumber \\
    &- \frac{2B}{1 + B} \, \frac{N-1}{N+1} \, {}_2F_1\bigg(1;- \frac{N}{2}+\frac{3}{2} ;\frac{N}{2} + \frac{3}{2};\frac{B - 1}{B + 1}\bigg)\bigg] \\
    \label{evenS}
    &S_\text{even} (N,B) = B +  \frac{(1 - B^2)^{\frac{N}{2}}}{2^N}\binom{N}{\frac{N}{2}}\bigg[\frac{1-B}{1 + B}  \nonumber \\
    &- 2B \, \frac{1 - B}{(1 + B)^2} \,  \frac{N-2}{N+2}  \, {}_2F_1\bigg(1; -\frac{N}{2}+2;\frac{N}{2}+2;\frac{B-1}{B+1}\bigg)\bigg]\, .
\end{align}
\end{subequations}
\begin{figure*}[t]
\begin{centering}
\includegraphics[width=1.0\textwidth]{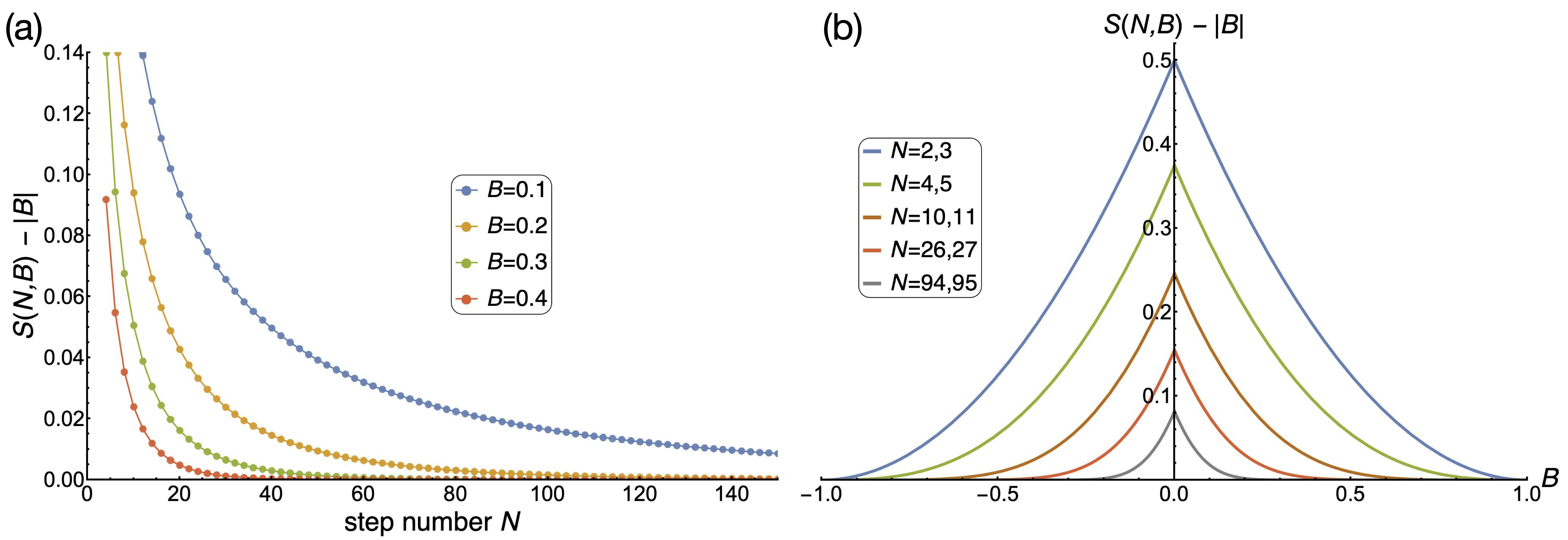}
\end{centering}
\vspace{-6mm}
\caption{Convergence properties of $S(N,B)$: We plot deviations of the survival probability from its asymptotic limit $|B|$ as a function of step number $N$ in panel (a), and as a function of the bias $B=p-q$ in panel (b). In panel (a) the points represent the analytic results ($N$ is discrete), which we have connected by a thin line for visual clarity. Heavily biased random walkers require few steps to approach the asymptotic limit, but small bias such as $B=0.1$ still require on the order of 100 steps to reach the asymptotic limit.}
\label{Deviations}
\end{figure*}These expressions hold for any step number ($N\geq1$) and any bias ($-1 < B < 1$). We were not able to locate the closed-form expression in Eq.~\ref{S} elsewhere in the literature. An alternate version of Eq.~\ref{S} expressed in terms of $p$ and $q$ is also provided in the Supplementary Material~\cite{SupplMat}. While we have provided both the even and odd step number formulae, only Eq.~\ref{evenS} is required to compute all survival probabilities.  This is because the survival probabilities for an even step number and its consecutive odd step number (e.g. $N=10$ and $N=11$) are the same, as illustrated in Fig.~1. Eq.~\ref{S} can be easily computed in standard software packages such as Mathematica and Matlab, in which the hypergeometric function is already implemented.  Our analytic result is compared with simulations in Fig.~\ref{Wedge}, showing excellent agreement.  The small bias behavior $S(N,B) \approx c(N) + d(N) \, B^2$ and the large bias behavior $S\approx |B|$ are also apparent.

\begin{figure}[b!]
\begin{centering}
\includegraphics[width=0.49\textwidth]{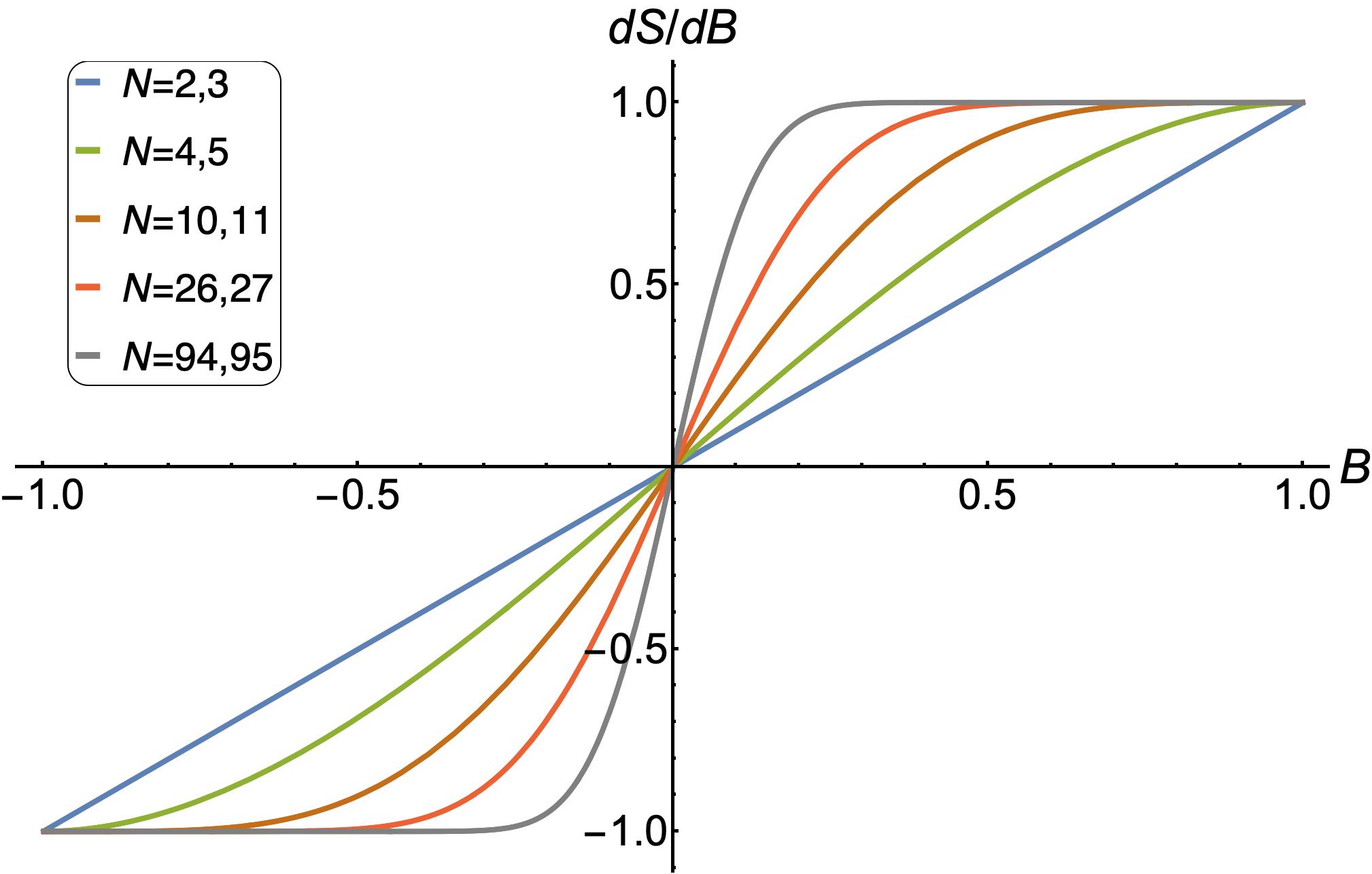}
\end{centering}
\vspace{-5mm}
\caption{Plot of the survival probability's derivative $dS/dB$ vs. bias $B$ for various step numbers $N$. The magnitude of $dS/dB$ saturates to $\pm 1$ as $|B|$ is increased, which corresponds to the asymptotic limit $S=|B|$.}
\label{BiasDeriv}
\end{figure}
Having an explicit expression of the survival probability as in Eq.~\ref{S} allows one to study its convergence properties with respect to increasing $N$ or bias $B$. To illustrate, in Fig.~\ref{Deviations} we plot the deviations $S(N,B)-|B|$ of the survival probability from its asymptotic limit, as a function of $N$ in panel (a) and of $B$ in panel (b).  Heavily biased walkers quickly approach the asymptotic limit, within $\sim \! \!10^0 - 10^1$ steps.  However, walks with a smaller bias such as $B\approx 0.1$ still require on the order of $10^2$ steps to reach the asymptotic limit.  

To examine the change in survival probabilities as the bias is continuously modulated, an explicit expression for $dS/dB$ can be obtained from Eq.~\ref{S}.  The following derivative property of the Gauss hypergeometric function~\cite{Hypergeometric}:
\begin{equation}
\frac{d }{dz} \ {}_2F_1(a,b;c;z) = \frac{a b}{c} \ {}_2F_1(a+1,b+1;c+1;z) 
\end{equation}
is used to produce Fig.~\ref{BiasDeriv} displaying $dS/dB$ (see Supplementary Material~\cite{SupplMat} for explicit formulae). The slope of a survival probability curve for a given $N$ becomes progressively larger in magnitude as $|B|$ increases, until the asymptotic behavior $dS/dB = \pm 1$ is reached (corresponding to $S = |B|$).

\begin{figure}[b!]
\begin{centering}
\includegraphics[width=0.49\textwidth]{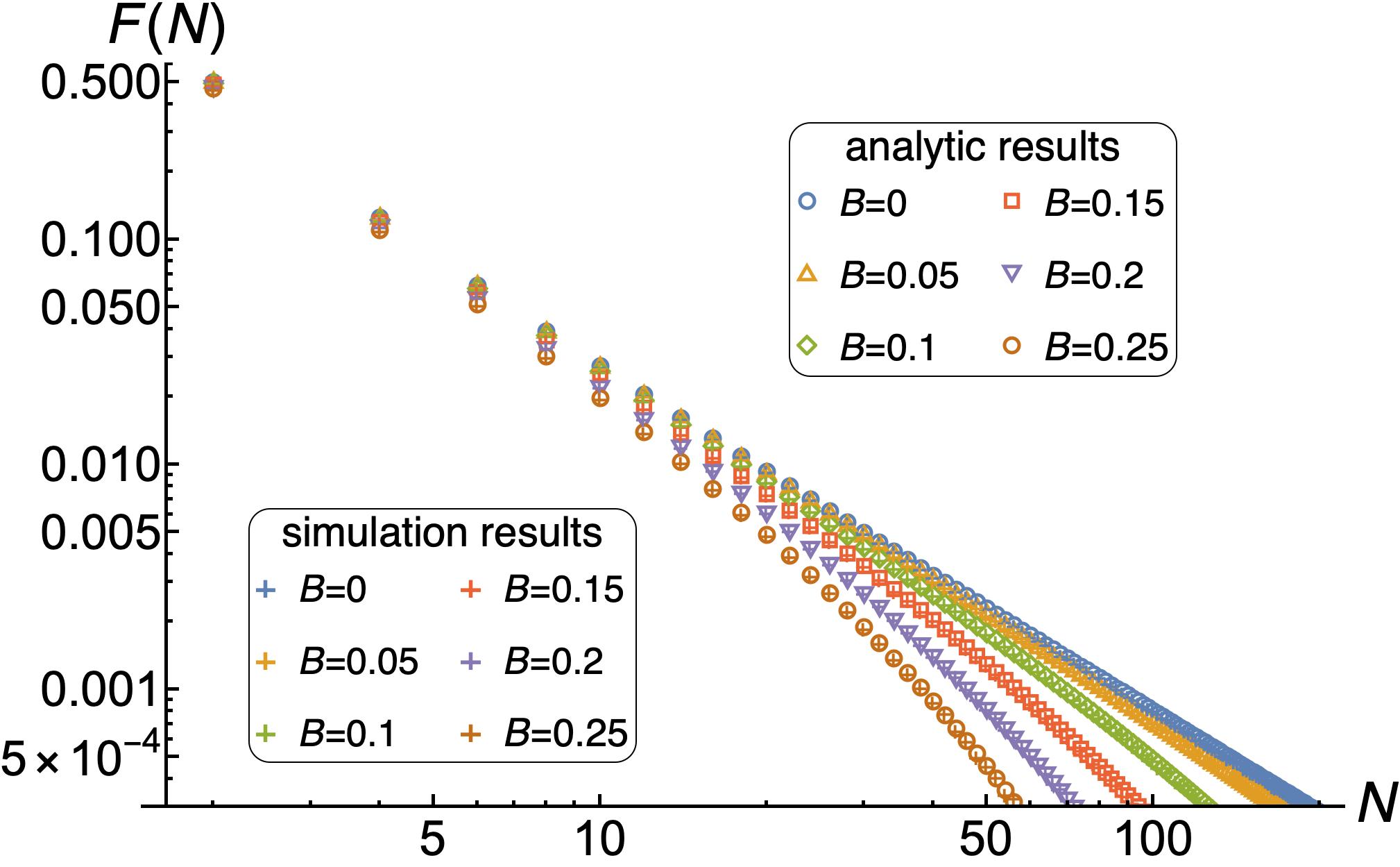}
\end{centering}
\vspace{-5mm}
\caption{The probability distribution $F(N)$ of first return occurring at (even) step number $N$: Shown are analytic results from Eq.~\!\ref{eq:Sdiff} (open markers) and simulation results (crosses) on a log-log plot.  The different colors correspond to six different values of the bias, ranging from $B=0$ to $B=0.25$. 
 The power law behavior $\sim \! N^{-3/2}$ for the symmetric case ($B=0$) is shown in blue.}
\label{FRTD}
\end{figure}
\begin{figure*}[t!]
\begin{centering}
\includegraphics[width=\textwidth]{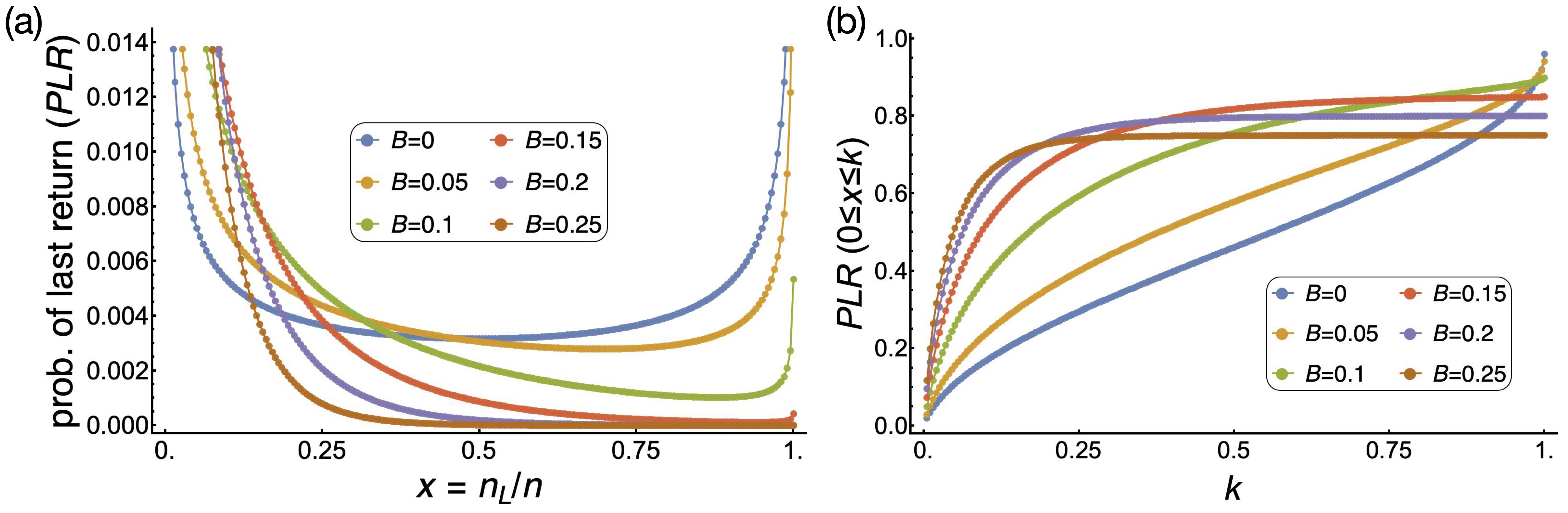}
\end{centering}
\vspace{-7mm}
\caption{Panel (a): Plot of the probability of last return (PLR) to the origin at step $2n_L$ for a walk consisting of $2n = 400$ steps. Here $x = n_L/n$ is the fraction of the walk already traversed. This plot illustrates deviations from the unbiased case (in blue). As the bias is increased, the peak near $x=1$ is diminished until it disappears entirely, and the PLR decreases monotonically throughout the walk. Panel (b): Plot of the cumulative of the PLR, i.e. the probability that the last return occurs in the first fraction $k$ of the walk.  Plotted here is $\text{PLR}(0 \leq x \leq k)$ for a total walk length of $2n = 400$ steps. For comparison, the arcsine behavior for a symmetric walk is shown in blue. The cumulative of the PLR saturates at $1-|B|$.}
\label{PLRfig}
\end{figure*}

From the survival probability, one can also easily obtain the probability distribution $F(N)$ of first return occurring at step $N$.  The survival probability is the complement of its cumulative: $\sum_{i=1}^{N} F(i)$ is the probability of return at step $N$, and it follows that
\begin{equation}
S(N) = 1 - \sum_{i=1}^{N} F(i) \, .
\end{equation}
We can thus determine $F(N)$ from the difference between survival probabilities at consecutive step numbers, denoted here by $i$ and $i+1$: 
\begin{equation}
    F(i+1) = -(S(i +1) - S(i)).
    \label{eq:Sdiff}
\end{equation} 
Because the survival probabilities for an even step number $i$ and its consecutive odd step number $i+1$ are the same, $F(i+1)$ is simply zero at odd step numbers.  This is in agreement with the fact that return can occur only at even step numbers.  Fig.~\ref{FRTD} compares the analytic results for even $N$ to simulations (see Supplementary Material~\cite{SupplMat} for $F(N)$ formulae). In the case of a symmetric walk ($B=0$), one recovers the power law behavior $\sim \! N^{-3/2}$, as shown in blue in Fig.~\ref{FRTD}.  

We now turn to the probability of last return (PLR). For a symmetric walk, the cumulative of the PLR yields the well-known arcsine law~\cite{Levy,Feller} which describes three quantities: (i) the proportion of time the walker spends on the positive half-line; (ii) the distribution of the last time the walker's position changes sign; and (iii) the distribution of times at which the walker's position reaches its maximum. Here we focus on the second interpretation and show how the arcsine law is modified once bias is introduced.  For a symmetric walk, the PLR is maximal at the beginning and end of the walk.  However we will show that for asymmetric walks, the PLR decays monotonically throughout the walk beyond a critical bias. 

Since returns to the origin occur only at even step numbers, let us denote the step number of last return as $2n_L$, and the entire length of the walk as $2n$.  The PLR is then the product of two probabilities: the probability to be at the origin at step $2n_L$, and the probability of surviving thereafter for $2n-2n_L$ steps.  Defining $x = n_L/n$ as the fraction of the walk completed at last return, we have
\begin{align}
    &\text{PLR($x, B$)} = \left(\frac{1 - B^2}{4}\right)^{n x} \binom{2nx}{nx} \, S_\text{even}(2n(1 - x),B)\nonumber\\
    &= \! \left[\frac{1 - B^2}{4}\right]^{n x} \binom{2nx}{nx} \biggl\{B +  \frac{(1 - B^2)^{n(1-x)}}{4^{n(1-x)}}\binom{2n(1-x)}{n(1-x)} \nonumber\\
    &\hspace{5mm} \times \bigg[\frac{1-B}{1 + B} - 2B \frac{(1 - B)}{(1 + B)^2}  \frac{n(1-x)-1}{n(1-x)+1} \nonumber\\
    &\hspace{5mm} \times {}_2F_1\bigg(1; 2-n(1-x);2+n(1-x);\frac{B-1}{B + 1}\bigg)\bigg]\biggl\} \, .
    \label{PLR}
\end{align}
Note that for the case $n_L = n$, the PLR is simply given by the probability to be at the origin at step $2n$: $\left(\frac{1 - B^2}{4}\right)^{n} \binom{2n}{n}$. We can then sum Eq.~\ref{PLR} for all steps up to $2n_L$ to obtain the cumulative of the PLR.  This yields the probability that the last return occurs sometime in the first fraction $n_L/n$ of the walk.
\begin{figure}[b!]
\begin{centering}
\includegraphics[width=0.49\textwidth]{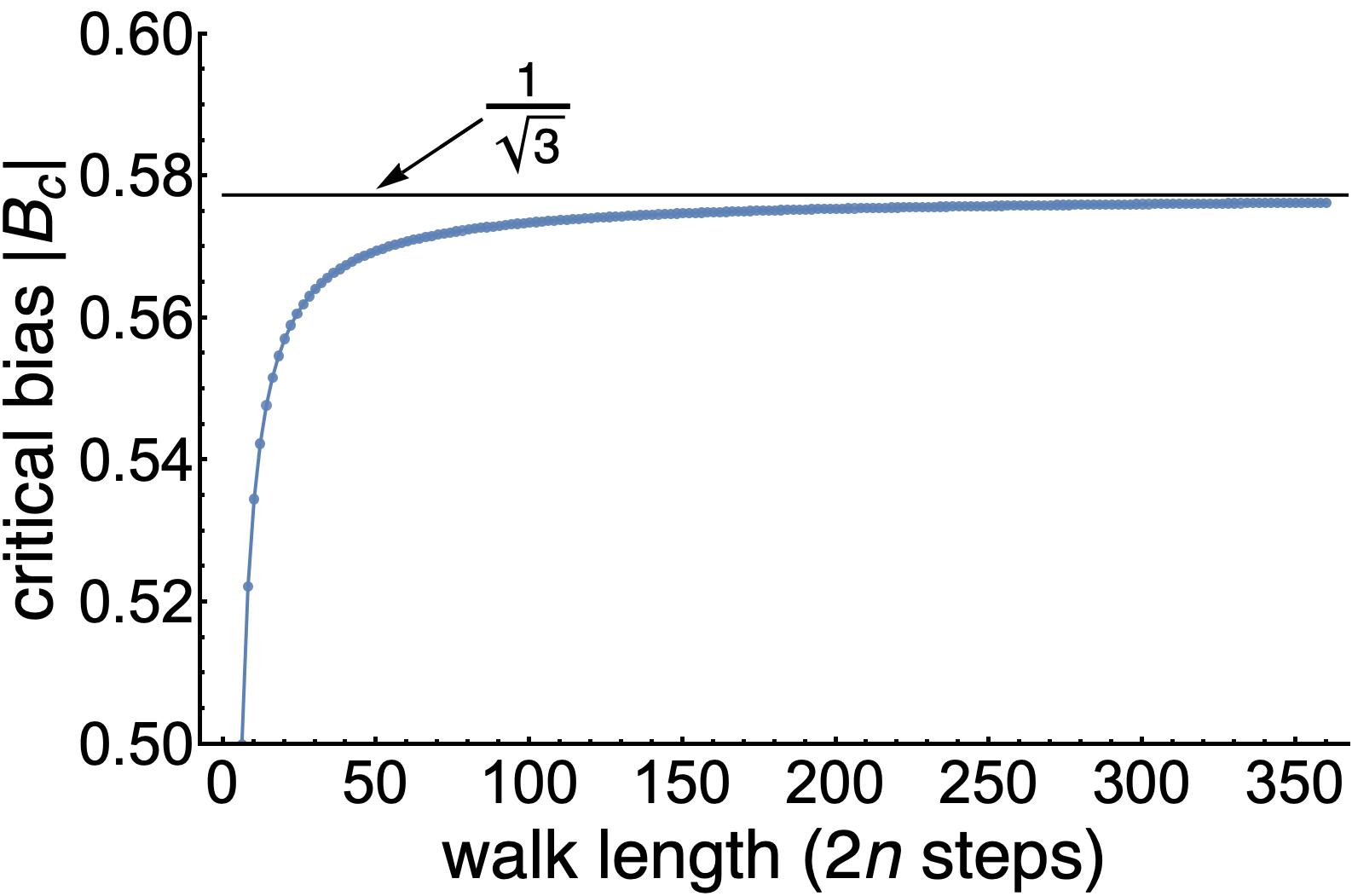}
\end{centering}
\vspace{-5mm}
\caption{The critical bias $|B_c| = \sqrt{(n-1)/(3n-1)}$ above which the probability of last return (PLR) decreases monotonically throughout a walk composed of $2n$ steps. This property of finite random walks was not noticed in previous works, as most expressions for the PLR are developed for infinitely long walks. Note that in the limit of large step number, the critical bias saturates at $1/\sqrt{3}\approx 0.577$.}
\label{Bc}
\end{figure}

Panel (a) of Fig.~\ref{PLRfig} illustrates how nonzero bias breaks the PLR symmetry of the unbiased case.  It also reveals that beyond a critical bias, the peak at $x=1$ disappears entirely and the PLR decays monotonically. The exact expression for the PLR here is essential: Such behavior is not evident from the large step number (long time) limit where Stirling's approximation is invoked. Note that the latter leads to a PLR of the form $\sim 1/\sqrt{x(1-x)}$, which yields a peak near $x=1$. Finally, in panel (b) we plot the cumulative, i.e. the probability of last return PLR($0 \leq x \leq k$) occurring sometime in the first fraction $k$ of a walk.  The arcsine law, which behaves as $2 \arcsin(\sqrt{k})/\pi$, is evident for zero bias (blue curve), and breaks down as the bias is increased.  One can also see that the cumulative of the PLR saturates to $1-|B|$ as $x \rightarrow 1$.

Finally, we determine the critical bias beyond which the PLR decreases monotonically (and thus the peak near $x=1$ disappears entirely) using Eq.~\ref{PLR}.  To do so, we examine when the difference in the PLR at penultimate and final steps of the walk, $\text{PLR}(1-1/n,B)-\text{PLR}(1,B)$ changes sign.  That is, we require $\text{PLR}(1,B)<\text{PLR}(1-1/n,B)$.  This leads to the condition:
\begin{equation}
    \left[\frac{1 - B^2}{4}\right]^{n} \binom{2n}{n} < \left[\frac{1 - B^2}{4}\right]^{n-1} \binom{2n-2}{n-1} \,  S_\text{even}(2,B)
\end{equation}
which simplifies to $(1-B^{2})(1-\frac{1}{2n}) < S_\text{even}(2,B)$.  Because $S_\text{even}(2,B) = (1+B^{2})/2$, the inequality can be reduced to the following (details provided in the Supplementary Material~\cite{SupplMat}): 
\begin{equation}
    |B| > \sqrt{\frac{n-1}{3n-1}} \, .
\end{equation} Hence we define the critical bias by $|B_c| = \sqrt{\frac{n-1}{3n-1}}$, which is plotted in Fig.~\ref{Bc}. Above this critical bias, a walk of total step number $2n$ will have a monotonically decreasing PLR.  Note that in the limit $n \rightarrow \infty$, the critical bias saturates at $1/\sqrt{3} \approx 0.577$.  Thus, there exists a critical bias of $1/\sqrt{3}$, beyond which random walks of any length will display a monotonically decreasing PLR.

In summary, our work fills a gap in the literature on biased random walks by deriving exact, closed-form expressions for recurrence/survival probabilities at arbitrary step number. Previous works on biased random walks neglected this in favor of studying the asymptotic behavior in the limit of large $N$. Our results will be useful for processes involving random walks of intermediate length.  Molecular motors like dynein and kinesin execute such walks along microtubules~\cite{Wordeman}, and can take over a hundred steps without detaching~\cite{Gennerich}. RNA polymerases may also be considered as executing biased random walks along mRNA~\cite{Roldan}. Such walks are biased in the forward direction: Dynein, for example, backtracks approximately 10 to 20\% of the time~\cite{Lecompte,Vale}, corresponding to a bias of $\approx \! 0.6$ to $0.8$. Intriguingly, these values remain above the critical bias value of $0.573$ for a typical walk length of $\sim$100 steps.

\acknowledgments
This work is in part supported by funding from New York University.

\end{document}